\documentclass[%
reprint,
superscriptaddress,
 amsmath,amssymb,
 aps,
pra,
]{revtex4-2}

\usepackage{graphicx}
\usepackage{dcolumn}
\usepackage{bm}

\begin{document}

\preprint{APS/123-QED}

\title{Characterization of cell division control strategies through continuous rate models}

\author{Cesar Nieto-Acuna}
 \email{ca.nieto13@uniandes.edu.co}
\affiliation{ 
Department of Physics, Universidad de los Andes, Bogot\'a, Colombia. South America\\
}

 \author{Juan Arias-Castro}%
 \affiliation{ 
Department of Physics, Universidad de los Andes, Bogot\'a, Colombia. South America\\
}
 \affiliation{Department of Systems Biology, Harvard Medical School, Boston, Massachusetts 02115, USA.\\}

 \author{Carlos Sanchez-Isaza}%
 \affiliation{ 
Department of Physics, Universidad de los Andes, Bogot\'a, Colombia. South America\\
}
 \affiliation{Department of Systems Biology, Harvard Medical School, Boston, Massachusetts 02115, USA.\\}

\author{Cesar Vargas-Garcia}%
\affiliation{%
Department of Mathematics and Engineering, Fundaci\'on universitaria Konrad Lorenz, Bogota, Colombia, South America\\
}%

 \author{Juan Manuel Pedraza}%
 \email{jmpedraza@uniandes.edu.co}
\affiliation{ 
Department of Physics, Universidad de los Andes, Bogot\'a, Colombia. South America\\
}%
\date{\today}

\begin{abstract}
Recent experiments have supported the \textit{Adder} model for \textit{E. coli} division control. This model posits that bacteria grow, on average, a fixed size before division.  It also predicts decorrelation between the noise in the added size and the size at birth. Here we use new experiments and a theoretical approach based on continuous rate models to explore deviations from the \textit{adder} strategy, specifically, the division control of \textit{E. coli} growing with glycerol as carbon source. In this medium, the division strategy is \textit{sizer-like}, which means that the added size decreases with the size at birth. We found that in a \textit{sizer-like} strategy the mean added size decreases with the size at birth while the noise in added size increases. We discuss possible molecular mechanisms underlying this strategy, and propose a general model that encompasses the different division strategies.
\end{abstract}

\maketitle
Bacterial homeostasis, the control of cell size distribution over a population of cells, has been extensively studied\cite{wang2010robust,marshall2012determines}. Determining the underlying mechanisms is important not only for a fundamental understanding of cell growth, but also because most forms of signaling inside cells depend on concentrations\cite{camilli2006bacterial}, which depend on cell volume, which in turn will fluctuate in a manner that depends strongly on the timing of cell division and its variability\cite{modi2017analysis}. Therefore, having an accurate stochastic model of cell division is paramount for predicting phenotypic variability and controlling intra-cellular circuits.

Experimental techniques have enabled high throughput measurements of cell growth dynamics, allowing the study of cell division strategies not only in bacteria such as \textit{Escherichia coli}\cite{campos2014constant}, \textit{Bacillus subtilis}\cite{taheri2015cell}, \textit{Caulobacter crescentus}\cite{iyer2014scaling} and \textit{Pseudomonas aeruginosa}\cite{deforet2015cell} but also yeast like \textit{Saccaromyces Cervisiae}  \cite{chandler2017adder} and \textit{Schizosaccharomyces pombe}\cite{nobs2014long}, and archea \cite{eun2018archaeal}, among others.

Division strategies, this is, how bacteria decide when to split into two descendant bacteria, can be classified on three main paradigms, each with different possible underlying mechanisms:  one is the \textit{timer} strategy, in which a cell waits a fixed time, on average, and then divides. A second paradigm is the \textit{adder}, in which a cell attempts to add a fixed size, on average, before dividing \cite{sauls2016adder}. The third is the \textit{sizer}, in which a cell grows until it reaches a certain volume\cite{facchetti2017controlling}.  

These strategies can be distinguished experimentally through measurements of added size vs. size at birth. An \textit{adder} would have a constant added size by definition, whereas a \textit{sizer} would produce a slope of -1, given the inversely proportional relationship between birth size and remaining growth needed to reach the desired fixed size. On the other hand \textit{timer} strategy slope is 1. This has a fundamental problem: is unable to produce stable cell size distributions when cells grow exponentially \cite{vargas2016conditions} and thus is not usually found in this kind of cells. This is not a problem for the \textit{adder} and \textit{sizer} models.

Biological systems can of course be more complicated, incorporating multiple controls that work in tandem or activate under different conditions. This leads to the phenomenological definition of \textit{sizer-like} \cite{vargas2016conditions} and \textit{timer-like} mechanisms, based upon the slope of added size vs. size at birth. Microorganisms like yeast\cite{facchetti2017controlling}, slow-growing  \textit{E. coli} cells\cite{wallden2016synchronization} and mycobacteria growing in sub-optimal growth media\cite{priestman2017mycobacteria} have been suggested as examples of \textit{sizer-like} behavior. 

Despite recent proposals \cite{si2018mechanistic,patterson2019noisy,witz2019initiation}, we still lack a mechanistic understanding of the biochemical mechanisms behind division control and how they depend on environmental conditions. One approach is to find the genes involved through traditional mutation assays\cite{sekar2018synthesis}, but another is to obtain a mechanistic model whose behavior matches experiments, and use it as a guide for which kinds of molecules to look for. Recent attempts at a mechanistic explanation include a threshold on the number of some precursor of division in the cell\cite{vargas2018elucidating,ghusinga2016mechanistic}.

The main idea behind this proposal is the modeling the cell decision through continous rate models (CRMs). These models consider not just discrete division events, but the continuous cell cycle. They specify the splitting rate function (SRF)\cite{taheri2015cell}, the instantaneous division rate, as a function of physiological parameters such as the current size, size at birth, growth rate, or the time since last division. Currently, the main problem on CRM is that it is not obvious
\textit{a priori} how to parametrize the SRF\cite{ho2018modeling}.

\begin{figure}
  \centering
  \includegraphics[width=0.45\textwidth]{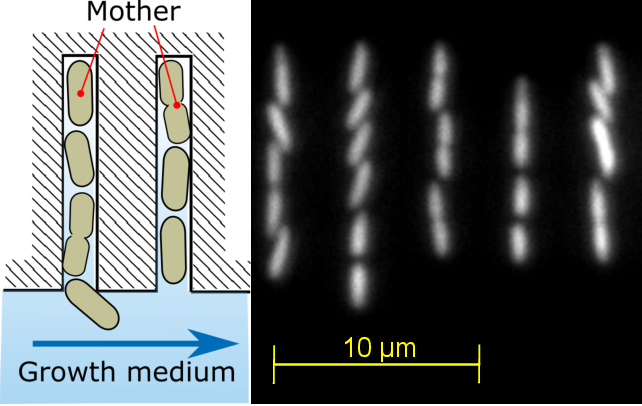}
  \caption{Left) schematic representation of the \textit{mother machine} micro-fluid. Right) Actual image in fluorescence channel of bacteria inside the growth trenches.}
  \label{fig:mm}
\end{figure}

To study division control, we use dynamic tracking of \textit{E. coli} cells growing in different media in a \textit{Mother Machine} microfluidic device\cite{taheri2015cell} (FIG. \ref{fig:mm}). This device enables the imprisonment of cells for measuring their size, growth and gene expression for hundreds of cell lineages over many generations while allowing continuous medium infusion to maintain balanced growth. We observe both \textit{adder} and \textit{sizer-like} behavior depending on the media.

We propose a biophysical model based on \cite{vargas2018elucidating}, consisting on a CRM with a SRF which does not depend linearly on the cell size but on a power of the size. Depending on the power, all the division paradigms can be modeled. We then refine our model by comparison with experiments, looking not only at the added size vs. size at birth but also at the noise in the added size. We also show how measuring population-wide dynamics on thousands of individual cells allow the use of noise characteristics as additional tools to distinguish between possible models.

\section{Theoretical considerations} 

We assume that each cell-cycle, i.e. the growth between a bacterial division and the next, can be modeled as exponential growth by the system of equations (\ref{law}):
\begin{equation}\label{law}
\dot{s}=\mu s\quad\dot{\tau}=1,
\end{equation}
with $s$ the cell size. $\mu$ is the growth-rate and $\tau$ is the time elapsed since the previous division, which is reset to 0 after every splitting event. In our experiments, cell length is used as a proxy of the cell size because cell width is mostly constant\cite{taheri2015cell} and measurements of the area have higher errors introduced by the small width.

Given a cell cycle time $\tau$, the probability of division during the time interval $(\tau,\tau + d\tau)$ is  described by the current division rate $h$\cite{tyson1986sloppy}. Here, we call $h$ the \textit{splitting rate function} (SRF) following notation used in other studies\cite{taheri2015cell,campos2014constant}. By this definition, it can be shown (see S.M.) that $h$ generates the cumulative distribution function (CDF) $P(\tau|s_b)$ of division at cell cycle time $\tau$ given the size at birth $s_b$:
\begin{equation}\label{srfdef}
h(s;s_b)=-\frac{d}{d\tau}\ln\left(1-P(\tau|s_b)\right).
\end{equation}
Thus, if $h$ can be obtained as a function of $\tau$, the integration of (\ref{srfdef}) can give us the distribution of division times.

With a SRF proportional to a power ($\lambda$) of the current cell size ($s$), we have, explicitly:
\begin{equation}\label{powerlaw}
h(s;s_b)\equiv ks^{\lambda}=ks^\lambda_b\exp(\lambda\mu\tau).
\end{equation}
After integration of (\ref{powerlaw}) in (\ref{srfdef}) (see S.M.), the probability distribution for cell splitting at size $s_d$ given its newborn size $s_b$ is:
\begin{equation}\label{analdist}
\rho(s_d|s_b)=\frac{k}{\mu}s_d^{\lambda-1}\exp\left(-\frac{k}{\mu\lambda}(s_d^\lambda-s_b^\lambda)\right)\theta(s_d-s_b)
\end{equation}
with $\theta(x)$ the Heaviside step function\cite{arfken1985mathematical}. From (\ref{analdist}) every $m$-th moment can be calculated, resulting in:
\begin{equation}\label{moments}
\mathbb{E}[s^m_d|s_b]=\exp\left(\frac{k}{\mu\lambda}s_b^\lambda\right)\left(\frac{\mu\lambda}{k}\right)^{\frac{m}{\lambda}}\Gamma\left(1+\frac{m}{\lambda},{\frac{k}{\mu\lambda}s_b^\lambda}\right)
\end{equation} 

where $\Gamma(a,z)=\int_z^\infty t^{a-1}e^{-t}dt$, the incomplete gamma function.

\begin{figure*}
  \centering
  \includegraphics[width=0.98\textwidth]{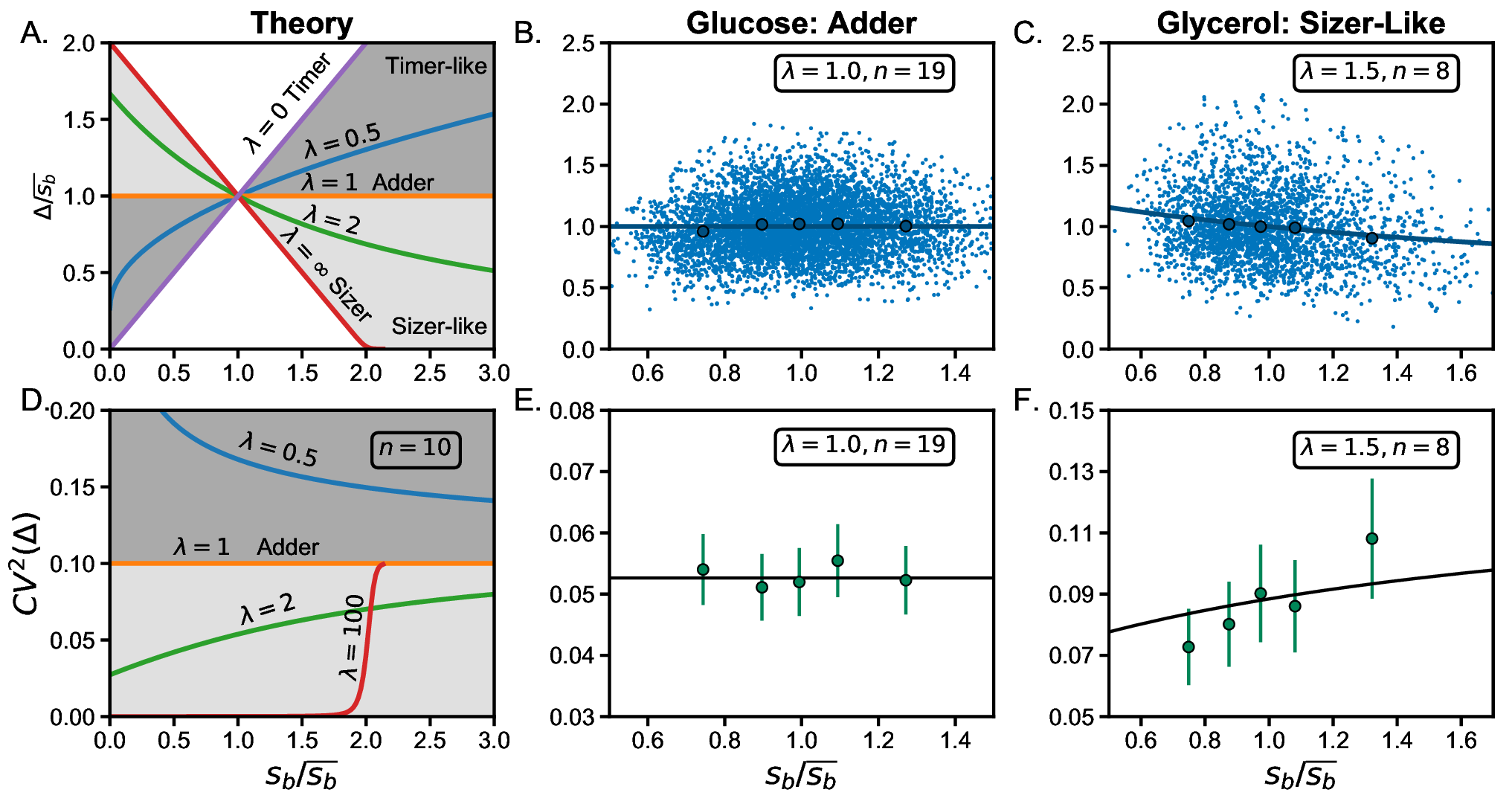}
  \caption{Comparison between the model predictions and measurements of cell division. A. General results for the added size ($\Delta$) as function of the newborn size ($s_b$). Division strategies (\textit{Timer}, \textit{Adder}, \textit{Sizer}) are shown as limit cases for the parameter $\lambda$ B. $\Delta$ vs $s_b$ for 5915 cell cycles of \textit{E. coli} growing with Glucose as carbon source. C.  $\Delta$ vs $s_b$ for 2740 cell cycles of \textit{E. coli} growing with Glycerol as carbon source. D. General results for $CV_\Delta^2$ vs $s_b$ for different division strategies. E. $CV_\Delta^2$ vs $s_b$ for \textit{E. coli} growing in Glucose F. $CV_\Delta^2$ vs $s_b$ for \textit{E. coli} growing in Glycerol. }
  \label{fig:main}
\end{figure*}

The expected added size $\mathbb{E}[\Delta]$ and the noise in added size $CV^2_\Delta$ can be obtained using:
\begin{eqnarray}\label{eq:momentis}
&&\mathbb{E}[\Delta|s_b]=\mathbb{E}[s_d|s_b]-s_b\nonumber\\ 
&&CV^2_\Delta(s_b)=\frac{\mathbb{E}[s^2_d|s_b]-(\mathbb{E}[s_d|s_b])^2}{\left(\mathbb{E}[s_d|s_b]-s_b\right)^2}.
\end{eqnarray}
The dependence of $\mathbb{E}[\Delta]$ on $s_b$ is shown in FIG. \ref{fig:main}.A. where the three main division strategies can be distinguished by their corresponding slope. They are: the perfect \textit{timer} strategy which is obtained when $\lambda\rightarrow 0$, the \textit{adder} when $\lambda\rightarrow 1$ and the perfect \textit{sizer} when $\lambda\rightarrow \infty$. Intermediate strategies are naturally obtained for intermediate $\lambda$. \textit{Timer-like} control is obtained when $0<\lambda<1$ and \textit{sizer-like} control when $1<\lambda<\infty$.

The typical size $\bar{s_b}$, as is shown in FIG. \ref{fig:main}.A., is the average cell size at birth. Theoretically, this size satisfies:
\begin{equation}\label{autoconsist}
\mathbb{E}[s_d|s_b=\bar{s_b}]=2\bar{s_b}.
\end{equation}

A closed-form expression for this $\bar{s_b}$ is not obtained, but it is possible to find it numerically using root-finding algorithms\cite{arfken1985mathematical} using (\ref{moments}). In FIG. \ref{fig:main}.A., the added size and the size at birth are normalized by $\bar{s_b}$ 

As it was pointed out in \cite{vargas2018elucidating} and found in \cite{si2018mechanistic}, cell division might require the completion of not one but $n$ events. In our data we see evidence indicating that the noise is far too low for a single step process (see next section). Therefore, (\ref{eq:momentis}) has to be rewritten to take into account $n$ successive but otherwise independent events each with a rate of occurrence $nh$ (see S.M.). The resulting distribution is the convolution of $n$ PDFs each following the equation (\ref{analdist}). It can be checked that $\mathbb{E}[s_d|s_b]$ does not change appreciably but $CV^2_\Delta$ is reduced:
\begin{equation}\label{eq:noiserep}
CV_n^2=\frac{1}{n}CV_1^2,
\end{equation}
where $CV_n^2$ is $CV^2_\Delta$ for a n-steps mechanism while $CV_1^2$ is the noise for a single step strategy. The relationship $CV_n^2$ vs $s_b$ is plotted in FIG \ref{fig:main}.D, where each strategy shows a different behavior with $s_b$. While $CV_n^2$ vs $s_b$ decreases for the \textit{timer-like} strategy, it increases for the \textit{sizer-like} strategy and is constant for the \textit{adder}.

\section{Experimental Approach}
To check the validity of these predictions, we performed time-lapse microscopy to monitor long-term single-cell size dynamics in \textit{Escherichia coli} bacteria growing in a \textit{mother machine} \cite{taheri2015cell} micro-fluidic device. This device consists on a series of small channels (growth trenches), oriented orthogonal to a larger channel through which growth medium (liquid) is passed at a constant rate (FIG. \ref{fig:mm}). This constant flow results in diffusion of fresh medium into the growth channels as well as removal of cells as they emerge from the side trenches into the main channel. The cell at the end of the growth trench, distal to the main channel, is referred to as the \textit{mother} cell because the entire lineage along the growth trench corresponds to descendants of this cell (FIG. \ref{fig:mm}).

Using the \textit{mother machine}, we tracked bacterial growth and division dynamics in two different carbon sources: Glucose (\textit{adder} strategy) and Glycerol (\textit{sizer-like} strategy) and compared these observations with our prediction for both division strategies.

\subsection{Experimental analysis of \textit{adder} mechanism: \textit{E. coli} in Glucose}
In FIG. \ref{fig:main}.B, the relationship between the added volume $\Delta=s_d-s_b$ and the newborn size $s_b$ is plotted for 5915 cell cycles of \textit{E. coli} growing in minimal medium M9 with Glucose as the carbon source. Details of the experimental methods can be found in the supplemental material. As it was reported previously\cite{taheri2015cell,campos2014constant}, these cells exhibit an \textit{adder} strategy incorporating, on average, a fixed cell size every division. 

For noise analysis (FIG. \ref{fig:main}.E.), the data were split in five quantiles. The five points shown in the graph, one per quantile, correspond to the average variation of data $(\Delta_i-\langle\Delta\rangle_q)^2/\langle\Delta\rangle^2_q$ around each quantile average $\langle\Delta\rangle_q$. The error bars correspond to a 95\% confidence interval and the solid line is the predicted value using equations (\ref{eq:momentis}) and (\ref{eq:noiserep}). For this \textit{adder} strategy, we found that the noise in added size ($CV^2_\Delta$), as expected from \cite{vargas2018elucidating}, shows no correlation with the size at birth. 

The number of steps in the mechanism can be estimated from the average noise using equation (\ref{eq:noiserep}) and the result $CV_\Delta^2=1$ for a single-step process. We obtain $n \approx 19$. This is a lower bound because there are sources of noise that we do not take into account explicitly, such as global noise \cite{pedraza2005noise,eldar2010functional}, or intrinsic noise in the proteins involved in the division decision \cite{swain2002intrinsic}. This means the actual number of steps could be higher and what we see is the overall effect of the noise in the division system plus other sources of noise but reduced via the multi-step process. 

\subsection{Experimental analysis of \textit{sizer-like} mechanism: \textit{E. coli} in Glycerol} 
FIG. \ref{fig:main}. C.  shows the division mechanism of \textit{E. coli} cells in Glycerol as carbon source. The amount of cell cycles analyzed for this condition is slightly lower (2740). We found a negative slope in the graph of $\Delta$ vs $s_b$. Although this slope is small ($\approx -0.24$), our data is sufficient to discard the null-hypothesis (p-value$\thicksim10^{-19}$) and this dependence has been reported previously\cite{taheri2015cell}. Our model can describe this behavior using $\lambda=1.5$ in (\ref{moments}). 

With $\lambda=1.5$ determined from the added size average we can fit $n \approx 8$ determined from the total average CV. While noise in added size does not show a correlation with size at birth in the \textit{adder} strategy (FIG. \ref{fig:main}.E.), according to our model it becomes an increasing function of the newborn size for a \textit{sizer-like} strategy (FIG. \ref{fig:main}.F.).

\section{Discussion} 
Here, we propose a biophysical model that reproduces the \textit{sizer-like} behavior observed in recent experiments but allows for other possible division strategies, as seen for growth in different media. The idea behind this control mechanism relies on the definition of a splitting rate function dependent on a power of the size. The dependence on size has been suggested by previous observations \cite{jasani2019growth,patterson2019noisy}.

A power of the SRF different from 1 could reflect the complexities of the processes needed to start division. Whereas $\lambda=1$ could correspond to simple accumulation of a molecule (such as FtsZ), $\lambda\neq 1$ could reflect cooperativity or interaction between different molecules needed to start division. In fact, some studies have shown that FtsZ molecules may be activated by other enzymes\cite{sekar2018synthesis,mannik2018cell}, and not only FtsZ but other division proteins can be heavily regulated at several levels\cite{dewar2000control}, some of which are not yet well known.

As an example of a possible mechanism, let us assume, following \cite{sompayrac1973autorepressor}, that there is an auto-repressor that controls the concentration of a molecule that tracks the size of the cell. The production rate r of such molecule can be written as
\begin{equation}
r = \frac{r_0}{1 + (c/ \bar{c})^h }
\end{equation}
where $c$ is the concentration of the molecule, $\bar{c}$ is a target concentration set by the auto-repressor mechanism, $h$ is the Hill coefficient of the repression, and $k$ is the maximum rate at which the molecule is produced. Writing it in terms of cell size ($s$) and molecule count ($p$) we have:
\begin{equation}
r = \frac{r_0}{1 + (c/ \bar{c})^h } = r_0 \frac{(\bar{c} s)^h}{( (\bar{c} s)^h + p^h ).}
\end{equation}
Assuming that at any cycle the initial count $p >> \bar{c} s$, we could approximate $r$ as $r \approx k\left(s\right)^h$ which is our proposed SRF. Thus changing the Hill coefficient ($h$) of $r$ is equivalent to changing $\lambda$ in our SRF. The Hill coefficient can arise from multimer formation or cooperativity in target binding, suggesting possible mechanisms for the observed \textit{sizer} behavior. These types of non-linear effects can be seen not only in the production of the molecules but also in the polymerization process\cite{nielsen2006progressive,lau2003spatial}. It is worth noting that a repressor mechanism based on concentration (local molecule count) is equivalent to a division process where a molecule is diluted, as suggested in some models of cell division. 

This framework describes the processes behind the biochemical mechanisms underlying the division control in \textit{E. coli} using only two free parameters ($\lambda$ and $n$) and describing not only the classical \textit{adder} division strategy but in other types (from \textit{timer} to \textit{sizer}) used in a range of different microorganisms\cite{sauls2016adder}. Fluctuations in division control ($CV_\Delta^2$) are also explained by this model. Because it does not identify the specific molecules, it can apply to multiple organisms, and the small number of parameters allow it to work also as a phenomenological description in cases where the biological details are still unknown. 

 The applications of this framework are extensive. The relationship between SRF functions and cell size control strategies further enable the use of recently proposed frameworks for gene expression\cite{jkedrak2019exactly} and cell lineage\cite{garcia2019linking} analysis of experimental data from proliferating cell populations. This work can also be used in instances regarding more complex organisms, e. g., homeostasis of organelle content in fission yeast\cite{jajoo2016accurate} which exhibits \textit{sizer-like} dynamics. It also gives us hints about how some eukaryotes could show clear departure from \textit{adder} by actively modulating physiological memory\cite{jasani2019growth}.
Finally, by condensing the effect of the molecular details into two effective parameters it allows the separate study of how growth conditions change the division strategy from the study of the effects of growth dynamics on expression noise and phenotypic variability.

\section{Acknowledgments}
We thank the Paulsson lab at Harvard Medical School for making their laboratory available for some of our measurements and image analysis. We also thank COLCIENCIAS convocatoria para doctorados nacionales 647 and Vicedecanatura de investigacion de la Facultad de Ciencias at Universidad de los Andes for financial support.

\bibliography{mainpaper}
\clearpage

\setcounter{section}{0}
\setcounter{equation}{0}
\setcounter{figure}{0}
\setcounter{table}{0}
\setcounter{page}{1}
\begin{center}
\large \textbf{Supplemental Materials: Characterization of cell division control strategies through continuous rate models}
\end{center}

\renewcommand{\theequation}{S\arabic{equation}}
\renewcommand{\thefigure}{S\arabic{figure}}
\section{Methods}
\textbf{Strain} \textit All the strains used in this study are \textit{Escherichia coli} k-12 MG1655 background \cite{jensen1993escherichia}. 

\textbf{Plasmid construction} We obtained a plasmid with GFP-mut2 under either promoters pNac or pRpoD from Uri Alon’s Plasmid Library\cite{zaslaver2006comprehensive} 
 and we modified the plasmid to insert a constitutive RFP promoter, induced by RNA1, as a segmentation marker. As a backbone we used the pUA66 plasmid from Uri Alon's library\cite{zaslaver2006comprehensive} and linearized it with a BglII restriction enzyme -Thermo scientific. Then the insert, mCherryKate2 constitutive marker was amplified from strain DHL60 from the Paulsson Lab at Harvard Medical School using colony PCR. Final assembly performed using the Gibson assembly protocol by New England BioLabs. 

\textbf{Growth media} Defined media was used in all experiments. For {Escherichia coli}, we used M9 minimal media\cite{miller1972experiments} with different carbon sources, glucose or glycerol, as shown in table \ref{tab:s1table1} 

\begin{table}[h!]
  \begin{center}
    \caption{M9 medium details.}
    \label{tab:s1table1}
    \begin{tabular}{l|r} 
       \textbf{Components} & \textbf{Concentration} \\
		\hline
      Disodium Phosphate & 48 mM\\
      Monopotassium Phosphate & 122 mM\\
      Sodium Chloride & 8.6 mM\\
      Ammonium Chloride & 18.7 mM\\
      Magnesium Sulfate & 1 mM\\
      Calcium Chloride & 0.5 mM\\
      Glucose (Glycerol) &0.2\% \\
      Kanamycin & 25 $\mu$l/ml\\
      \hline
      BSA & 0.5 mg/ml\\  
      pluronic F108& 0.8 g/l\\    
    \end{tabular}
  \end{center}
\end{table}

\textbf{Cell preparation}. Before every time-lapse imaging, cells were picked from a single colony on an agar plate which was streaked no more than 7 days before use. The cells were inoculated into M9 with selection antibiotics; in our case kanamycin 25 ul/ml. After 12-18 hours at 37 $^o$C in a water bath shaker, cells were diluted 1,000-fold into 2 mL of the same defined medium as that used in microfluidic experiment. After shaking at 37$^o$C in a water bath untill reaching OD600 = 0.1-0.4, cells were diluted again 100- to 1,000-fold into the same medium and shaken at 37 $^o$C in a water bath untill OD600 = 0.2. The cell culture was then concentrated 10- to 100-fold and injected into a microfluidic \textit{mother machine} device via a micropipette with gel-loading tips. Moreover, 0.5 mg/ml BSA -Bovine serum albumin, Gemini Bio Products, CA-  was added to the fresh growth media to reduce the adhesion of cells to the surface of microfluidic channels. The media was then added to 60 mL plastic syringes -BD- and flowed using a syringe pump with a flow of 30 $\mu l/min$ for time-lapse imaging. All imaging experiments were conducted at 37 $^o$C in an environmental chamber. 

\textbf{Microfluidics}  \textit{Mother machine} microfluidic devices were used in this study to monitor single cell growth for 12-20 generations. Master molds, from each of which many PDMS microfluidic devices were cast, were fabricated using standard nanofabrication techniques (detailed protocols are available in “The Mother Machine Handbook” from the web site of Jun lab at http://jun.ucsd.edu and a video at http://www.youtube.com/watch?v=RGfb9XU5Oow), courtesy of the Paulsson lab.

PDMS was prepared from a Sylgard 184 Silicone Elastomer kit: polymer base and curing agent were mixed in a 10 to 1 ratio, air bubbles were purged from the mixture in a vacuum chamber, the degassed mixture was poured over the master, and the devices were cured for about 1h at 90$^o$C. Cured PDMS has a rubber-like consistency which allows devices to be peeled manually from the master mold. Devices were treated with Isopropyl Alcohol to remove residual uncured polymer from the PDMS matrix.

To bond the PDMS layers, the surfaces were exposed to oxygen plasma for 15 seconds at 30 watts in a Harrick Plasma system. Oxygen plasma makes exposed PDMS and glass reactive, so that covalent bonds form between surfaces brought into contact with one another. The seal between PDMS surfaces was established for 10 minutes at 65 $^o$C.

\textbf{Microscopy and image acquisition}. 
An inverted Nikon Eclipse Ti microscope equipped with Perfect Focus system, a 60x air objective lens (NA 0.95), a lumencor spectra X3 light engine and a an Andor Zyla 4.2 PLUS sCMOS camera were used for fluorescense imaging. The filter set used was the ET-Sedat Quad-band (8900, Chroma Technology Crop). The exposure time was set to 200 ms and the illumination intensity at 100\%. The time-lapse frequency was 15 min or 22 min.

\textbf{Image analysis}
Briefly, the segmentation was done using images from a bright, constitutively expressed RFP on a PRNA1 promoter. The rough trench boundaries were estimated, with the Otsu threshold method followed by erosion, opening and dilation of the mask. The bounding box of the found trenches was then used to find cells within. Then, on each of the trenches the cells were segmented using Niblack segmentation\cite{samorodova2016fast}. Cells joined by their poles (as indicated by objects with definite constrictions) were separated using the top 10\% brightest pixels of cells as seed for watershed. Spurious non-cell objects were rejected using their size, orientation and shape. Finally, the boundaries were refined using opening, thickening and active contours. The parameters chosen for each experiment required extensive testing and the segmentation was checked manually. We chose to follow only the cells at the closed end of the channel.

\textbf{Data Analysis}

Once the size of all the \textit{mothers} over time is obtained, cell cycles can be discriminated by finding the time when the cell size divides. This is done by a peak detector \cite{palshikar2009simple} which only considers high variations (up to 33\% the cell size) on the first order difference in adjacent size values.

\begin{figure}[h!]
  \centering
  \includegraphics[width=0.45\textwidth]{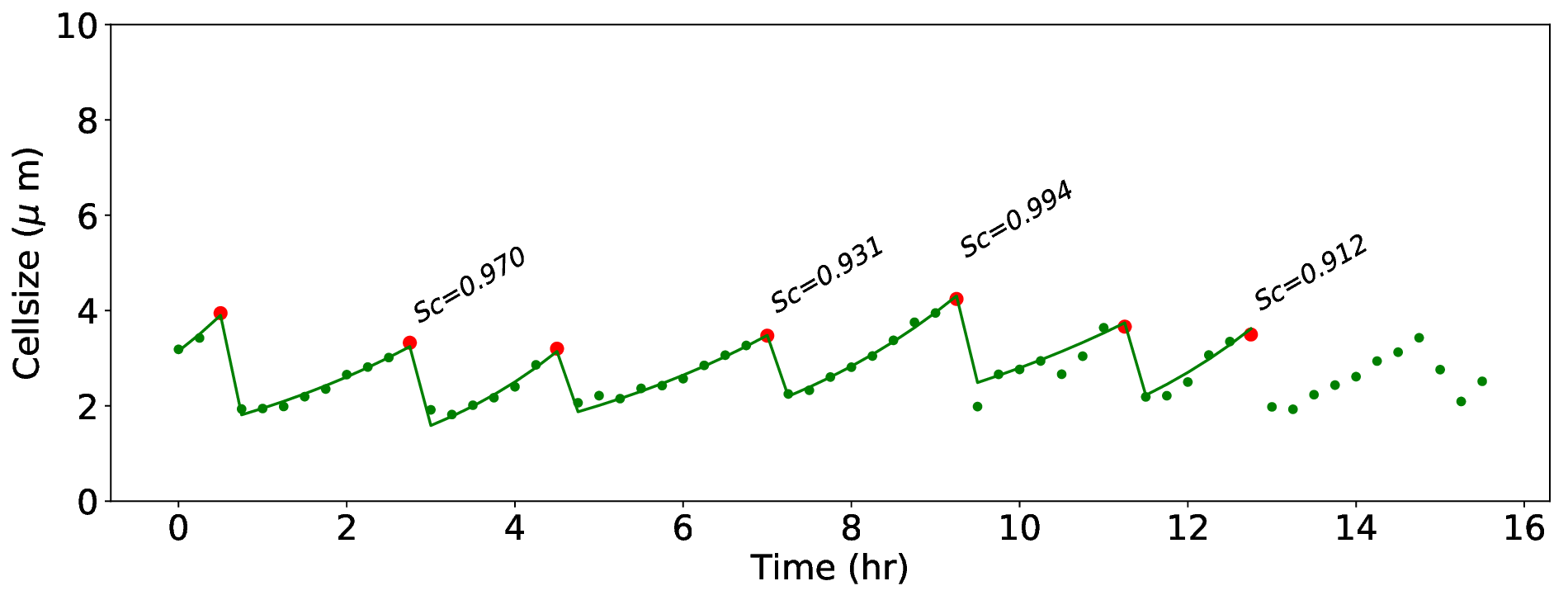}
  \caption{\textit{Mother} size dynamics in a typical time-lapse. The division times are obtained from the frames with high variations in the first order difference between adjacent size values (red dots). The points between divisions are used for an exponential fit (green line). The exponential parameter is taken as the growth rate. The score of the fit from an interval is shown over the peaks.}
  \label{fig:signalex}
\end{figure}

One example of the result of the signal processing made for a real experiment of cell growth is shown in FIG \ref{fig:signalex} where the peaks obtained with our peak detector are shown (red dots). Once the division times are obtained, the growth rate can be estimated by fitting the points corresponding to the cell cycle. With the supposition that bacterial cell size grows obeying an exponential growth law, the size s(t) along the time can be modeled by the equation:
\begin{equation}
S(t)=s(0)2^{\nu t}
\end{equation}

Where s(0) is the cell size once the measurement begins and $\nu$ is defined as the growth rate.

We use a Random sample consensus (RANSAC) \cite{fischler1981random} estimator for the fitting between the logarithm of the size to base 2 and the time during each cell cycle. The RANSAC algorithm detects the outliers and discards them. This is important in cases where there are errors in the segmentation and high deviations to the trend line appear. These errors usually occur via a bad segmentation in which two bacteria are taken as one. An example of how this fitting is made is shown in FIG. \ref{fig:signalex} where the score of the fitting for each cell cycle is shown over the peaks. 

These sources of error affect the determination of the growth-rate, resulting in many intervals where the exponential is a poor fit. There are also many intervals without obvious segmentation errors where the behavior is not exponential, and this depends on the medium: $\approx 48\%$ of the glycerol cell cycles satisfy $R^2>0.9$ for an exponential fit while $\approx 85\%$ glucose cells cycles do.
\section{Theoretical details}
\subsection{The Splitting rate function}
We define the splitting rate function (SRF) $h(s,s_b)$ as the local Poisson rate of cell division in cell age ($\tau_d$), i.e.
\begin{equation}\label{eq:srf}
    P(\tau_{d}\in(\tau,\tau+d\tau))=hd\tau
\end{equation} 
is the probability that a cell with size $s$ and newborn size $s_b$ divides within the time interval $\tau$, $\tau+d\tau$. 

Suppose that we want to compute the probability that a cell divides while it's growing from size $s$ to $s+ds$ given the size at birth $s_b$. Thus, we can start from the probability that it had not divided during the time interval $(0,\tau)$ and then divided during the time interval $(\tau,\tau+d\tau)$ at size $s(\tau)=s_d$. Given the SRF (\ref{eq:srf}), we have:
\begin{equation}\label{rho}
\rho(s_d|s_b)ds=\exp\left(-\int_0^{\tau} h(s(\tau'))d\tau'\right)h(s(\tau))d\tau
\end{equation}
Most \textit{adder} models consider this SRF as dependent only on the added size: $h(s(\tau),s_b )=h(s(\tau)-s_b )=h(\Delta(\tau))$\cite{taheri2015cell,tyson1986sloppy}. Here, we will show how an \textit{adder} mechanism can arise assuming a SRF dependent on the current size.
 
If we use
\begin{equation}
h(s)\exp\left[-\int_0^{\tau}h(s(\tau'))d\tau'\right]
   =-\frac{d}{d\tau}\exp\left[-\int_0^{\tau}h(s(\tau'))d\tau'\right]
\end{equation}
, we can rewrite the equation (\ref{rho}) as:
\begin{equation}
   \rho(s_d=s|s_b)\frac{ds}{d\tau}=-\frac{d}{d\tau}\exp\left[-\int_0^{\tau}h(s(\tau'))d\tau'\right]
\end{equation}
which, after integration, can be written as:
\begin{equation}
\int_0^\tau \rho(\tau_d=\tau'|s_b)d\tau' -1
=-\exp\left[-\int_0^{\tau(s_d,s_b)}h(s(\tau'))d\tau'\right]
\end{equation}

After some algebra, we obtain:

\begin{eqnarray}
\label{relationship}
h(s(\tau))&=&-\frac{d}{d\tau}\ln\left(1-\int_0^\tau\rho(\tau_d=\tau'|s_b)d\tau'\right)\nonumber\\
&=&-\frac{d}{d\tau}\left(\ln(1-P(\tau_d=\tau|s_b))\right)
\end{eqnarray}
where $\rho(\tau_d=\tau|s_b)d\tau$ is the probability that a cell splits during the time interval $(\tau,\tau+d\tau)$, given a size at birth $s_b$ and the growth rate $\mu$;
and $P(\tau_d=\tau|s(0)=s_b)$, or simply $P(\tau_d|s_b)$, is the cumulative distribution function (CDF) for this division probability. By the relationship (\ref{relationship}), given the SRF, the cumulative distribution can be estimated as follows.

\subsection{One-step \textit{adder}}

A SRF for a simple \textit{adder} could be one in which it is proportional to the current cell size ($s$). 
\begin{equation}\label{eq:adderans}
    h(s(\tau))=ks=ks_b\exp(\mu \tau)
\end{equation}
with $k$ some constant. Taking into account the exponential growth law, there is a relationship between this size ($s$) and the cell age ($\tau$):
\begin{equation}\label{eq:growthla}
\tau=\frac{1}{\mu}\ln\left(\frac{s}{s_b}\right)
\end{equation}.

With the SRF given in (\ref{eq:adderans}), (\ref{relationship}) can be solved for $\rho(\tau_d|s_b)$ as follows. We can start by solving for the CDF:
\begin{equation}
    P(\tau_d|s_b)=1-\exp\left(\frac{ks_d}{\mu}(1-e^{\mu\tau_d})\right)
\end{equation}
which gives a probability density function (PDF) of cell splitting at time $\tau=\tau_d$
\begin{eqnarray}
    \rho(\tau=\tau_d|s_b)&=&\frac{dP(\tau_d|s_b)}{d\tau}\nonumber\\
    &=&ks_b\exp\left(
    \mu\tau_d+
    \frac{ks_b}{\mu}
    (1-e^{\mu\tau_d})\right).
\end{eqnarray}

Using the relationship (\ref{eq:growthla}), we obtain:
\begin{eqnarray}
\rho(s_d|s_b)&=&\rho(\tau_d|s_b)\frac{d\tau}{ds}=\frac{k}{\mu}\exp\left(-\frac{k}{\mu}(s_d-s_b)\right)\nonumber\\&=&\frac{k}{\mu}\exp\left(-\frac{k}{\mu}\Delta\right)
\end{eqnarray}

If we define the added size $\Delta=s_d-s_b$, the distribution of this added size is given by:
\begin{eqnarray}\label{adderdist}
\rho(s=s_d|s_b)&=&\rho(s_d-s_b)=\rho(\Delta)\nonumber\\
&=&\frac{k}{\mu}\exp\left(-\frac{k}{\mu}\Delta\right)
\end{eqnarray}
This means that in this model the added size does not depend neither on the splitting time nor on the newborn size and is an exponentially distributed variable with mean
\begin{equation}
    E[\Delta]=\frac{\mu}{k}.
\end{equation}
\subsection{The multi-step \textit{Adder}}\label{multisect}
In the multi-step \textit{adder} strategy, the division depends on the occurrence of $n$ single-step \textit{adders} before division. In this model, the size at division can be written as
\begin{equation}
    s_d=s_b+\Delta
\end{equation}

where $\Delta=\sum_{i+1}^{n}\Delta_i$ and $\Delta_i$ is the added size at each of the of the $n_s$ steps required to divide. Each step can be taken as an exponential distributed variable with parameter $nk/\mu$.\\
Since the $\Delta_i$ are exponentially distributed and independent variables, the distribution of $\Delta$ is the convolution of $n_s$ exponential PDFs with parameter $nk/\mu$:
\begin{equation}\label{eq:multistep}
    \rho(\Delta)=\left(\frac{nk}{\mu}\right)^{n}\frac{\Delta^{(n-1)}}{(n-1)!}\exp\left(-\frac{nk}{\mu}\Delta\right)
\end{equation}

With this distribution, the mean value of the size at division $s_d$, given the size at birth $s_b$, is
\begin{equation}
    E[s_d|s_b]=s_b+E[\Delta]=s_b+\frac{\mu}{k}
\end{equation}
while other important parameters are:
\begin{eqnarray}
Var(\Delta)=\frac{1}{n}\left(\frac{\mu}{k}\right)^2
CV^2_\Delta=\frac{1}{n}\\
Skew=\frac{2}{\sqrt{n}}
\end{eqnarray}

By the relationship $\Delta(\tau)=s_b(e^{\mu\tau}-1)$, we can compute the PDF for the division times:
\begin{equation}
    \rho(\tau|s_b)=\mu s_b e^{\mu\tau}\left(\frac{nk}{\mu}\right)^n\frac{(\Delta(\tau))^{n-1}}{(n-1)!}\exp\left(\frac{-nk}{\mu}\Delta(\tau)\right)
\end{equation}

We can then infer the SRF using the relationship :
\begin{equation}
h(s(\tau),s_b)=\frac{\rho(\tau|s_b)}{1-\int_0^\tau\rho(\tau'|s_b)d\tau'}    
\end{equation}
\subsection{Single-step nonlinear mechanism}
Deviations from the adder strategy can arise from a non-linear SRF dependence on the size. Consider a SRF proportional to a power $\lambda$ of the size:
\begin{equation}
    h(s,s_b)=ks^{\lambda}=ks_b^\lambda e^{\lambda\mu\tau}
\end{equation}
where $s$ is a re-scaled size. 

Using a similar procedure to the previous section, we can be obtain:
\begin{eqnarray}
      P_\lambda(\tau_d|s_b)&=&1-\exp\left(\frac{-k}{\mu\lambda}s_b^\lambda\left(e^{\lambda\mu\tau}-1\right)\right)\\  
      \rho_\lambda(s_d|s_b)&=&\frac{k}{\mu}s^{\lambda-1}\exp\left(-\frac{k}{\mu\lambda}(s_d^\lambda-s_b^\lambda)\right)\theta(s_d-s_b)\nonumber\label{eq: distrnonlin}
\end{eqnarray}
Unlike (\ref{adderdist}), in (\ref{eq: distrnonlin}), the added volume is dependent on the birth size $s_b$.
Thus, the expected value of the size at division given the size at birth is given by:
\begin{eqnarray}
    E[s_d|s_b]&=\exp\left(\frac{k}{\mu\lambda}s_b^\lambda\right)\frac{k}{\mu}\int_{s_b}^\infty s_d^\lambda\exp\left(\frac{k}{\mu\lambda}s_d^\lambda\right)ds_d\nonumber &\\
    &=\exp\left(\frac{k}{\mu\lambda}s_b^\lambda\right)\left(\frac{\mu\lambda}{k}\right)^{\frac{1}{\lambda}}\Gamma\left(1+\frac{1}{\lambda},{\frac{k}{\mu\lambda}s_b^\lambda}\right)&
\end{eqnarray}

where $\Gamma(a,z)$ is the incomplete gamma function $\Gamma(a,z)=\int_z^\infty t^{a-1}e^{-t}dt .$

In general, the $\alpha$th distribution moment can be obtained as function of $s_b$:
\begin{equation}\label{eq: moments}
    E[s_d^\alpha|s_b]=\exp\left(\frac{k}{\mu\lambda}s_b^\lambda\right)\left(\frac{\mu\lambda}{k}\right)^{\frac{\alpha}{\lambda}}\Gamma\left(1+\frac{\alpha}{\lambda},{\frac{k}{\mu\lambda}s_b^\lambda}\right)
\end{equation}
Thus, we can obtain the variance in the added size given the size at birth $Var[\Delta|s_b]$

\begin{eqnarray}
&\rm{Var}[\Delta|s_b]=E[s_d^2|s_b]-(E[s_d|s_b])^2&\\
    &=\exp\left(\frac{k}{\mu\lambda}s_b^\lambda\right)\left(\frac{\mu\lambda}{k}\right)^{\frac{2}{\lambda}}\Gamma\left(1+\frac{2}{\lambda},{\frac{k}{\mu\lambda}s_b^\lambda}\right)-&\nonumber\\&\left(\exp\left(\frac{k}{\mu\lambda}s_b^\lambda\right)\left(\frac{\mu\lambda}{k}\right)^{\frac{1}{\lambda}}\Gamma\left(1+\frac{1}{\lambda},{\frac{k}{\mu\lambda}s_b^\lambda}\right)\right)^2&
\end{eqnarray}
Here we see that not only the mean value but the noise in added size (expressed by the Coefficient of variation $CV^2_\Delta=\rm{Var}[\Delta|s_b]/(E[\Delta|s_b])^2$), depends on the added size.

\subsection{Multistep non-linear mechanism}

Exponent $\lambda$ can define not only the slope on $\Delta$ vs $s_b$ but the noise signature $CV^2_\Delta$ vs $s_b$. However, the added size shows a noise lower than expected from a one step mechanism\cite{taheri2015cell}. As in section \ref{multisect}, we can propose a multistep-non linear mechanism. Following the same procedure shown in (\ref{eq:multistep}), we can use the analytic expression for all the moments in (\ref{eq: moments}) to obtain the moment generating function for the first moments of the multi-step mechanism.

Let us assume that the added size is the sum of $n$ events modeled by a non-linear SRF with parameter $nk$ and start with the power expansion of the moment generating function for a convolution of $n$ processes distributed as (\ref{eq: distrnonlin}). If $M(t|s_b)$ is the moment generating function for a single step division mechanism (\ref{eq: distrnonlin}), we can infer some of the moments of the convolution of $n$ steps with a non-linear SRF:
\begin{eqnarray}\label{eq:multistepnon}
       M_n(t)&=&(M(t|s_b))^{n}=\left(\sum_{k=0}^\infty \frac{E[s_d^k|s_b]t^n}{k!}\right)^n\nonumber\\
       &\approx& 1+nE[s_d|s_b]t\nonumber\\&&+\left[nE[s^2_d|s_b]+\frac{n(n-1)}{2}(E[s_d|s_b])^2\right]\frac{t^2}{2}
\end{eqnarray}

Taking the first two terms in the expansion, we obtain the analytic formulae:
\begin{small}
\begin{eqnarray}
    E_n[s_d|s_b]=n\exp\left(\frac{nk}{\mu\lambda}s_b^\lambda\right)\left(\frac{\mu\lambda}{nk}\right)^{\frac{1}{\lambda}}\Gamma\left(1+\frac{1}{\lambda},{\frac{nk}{\mu\lambda}s_b^\lambda}\right)\nonumber\\
    E_n[s^2_d|s_b]=n\exp\left(\frac{nk}{\mu\lambda}s_b^\lambda\right)\left(\frac{\mu\lambda}{nk}\right)^{\frac{2}{\lambda}}\Gamma\left(1+\frac{2}{\lambda},{\frac{nk}{\mu\lambda}s_b^\lambda}\right)+\nonumber
    \\n(n-1)\left(\exp\left(\frac{nk}{\mu\lambda}s_b^\lambda\right)\left(\frac{\mu\lambda}{nk}\right)^{\frac{1}{\lambda}}\Gamma\left(1+\frac{1}{\lambda},{\frac{nk}{\mu\lambda}s_b^\lambda}\right)\right)^2\nonumber
\end{eqnarray}
\end{small}
The noise signature for a multi-step mechanism $CV_n^2$ can be related with that of the single step one $CV_1^2$ by:
\begin{equation}
CV_n^2=\frac{1}{n}CV_1^2
\end{equation}
where $CV_1^2$ has parameter ($nk$) instead of ($k$).
\end{document}